\documentclass[apj]{emulateapj}
\usepackage{color}

\newcommand{\eg}{e.g., }
\newcommand{\ie}{i.e., }
\newcommand{\Msun}{M_{\odot}}
\newcommand{\kms}{km~s$^{-1}$}

\newcommand{\Nifs}{$^{56}$Ni}

\def\gsim{\mathrel{\rlap{\lower 4pt \hbox{\hskip 1pt $\sim$}}\raise 1pt
\hbox {$>$}}}
\def\lsim{\mathrel{\rlap{\lower 4pt \hbox{\hskip 1pt $\sim$}}\raise 1pt
\hbox {$<$}}}

\def\ion#1#2{{\rm #1}~{\sc #2}}

\shorttitle{Spectropolarimetry and Asphericity of SN 2007gr}
\shortauthors{Tanaka et al.}
 
\begin{document}

\title{
Optical Spectropolarimetry and Asphericity of 
Type I\lowercase{c} SN 2007\lowercase{gr}
\altaffilmark{1}}
\author{
Masaomi Tanaka \altaffilmark{2},
Koji S. Kawabata \altaffilmark{3},
Keiichi Maeda \altaffilmark{4},
Takashi Hattori \altaffilmark{5},
and
Ken'ichi Nomoto \altaffilmark{4,2}
}

\altaffiltext{1}{Based on data collected at Subaru Telescope, 
which is operated by the National Astronomical Observatory of Japan.}
\altaffiltext{2}{Department of Astronomy, Graduate School of Science, University of Tokyo, 7-3-1 Hongo, Bunkyo-ku, Tokyo 113-0033, Japan; mtanaka@astron.s.u-tokyo.ac.jp}
\altaffiltext{3}{Hiroshima Astrophysical Science Center, Hiroshima University, Higashi-Hiroshima, Hiroshima, Japan}
\altaffiltext{4}{Institute for the Physics and Mathematics of the
Universe, University of Tokyo, Kashiwa, Japan}
\altaffiltext{5}{Subaru Telescope, National Astronomical Observatory of
Japan, Hilo, HI}

\begin{abstract}
We present optical spectropolarimetric observations 
of Type Ic supernova (SN) 2007gr with Subaru telescope
at 21 days after the maximum brightness ($\sim 37$ days after the explosion).
Non-zero polarization as high as $\sim 3$ \% 
is observed at the absorption feature of \ion{Ca}{ii} IR triplet.
The polarization of the continuum light is $\sim 0.5$\%
if we estimate the interstellar polarization (ISP) component
assuming that the continuum polarization has a single 
polarization angle.
This suggests that the axis ratio of the SN photosphere projected to the sky 
is different from unity by $\sim 10$\%.
The polarization angle at the \ion{Ca}{ii} absorption is almost aligned 
to that of the continuum light.
These features may be understood by the model 
where a bipolar explosion with an oblate photosphere 
is viewed from the slightly off-axis direction 
and explosively synthesized Ca
near the polar region obscures the light originated around 
the minor axis of the SN photosphere.
Given the uncertainty of the ISP, however,
the polarization data could also be interpreted by 
the model with an almost spherically symmetric photosphere 
and a clumpy \ion{Ca}{ii} distribution.
\end{abstract}


\section{Introduction}
\label{sec:intro}

The explosion mechanism of core-collapse supernovae (SNe) is 
still under debate.
Recent theoretical studies show that 
the effects causing non-spherical explosion, 
such as magnetic field, rotation, and several kinds of instabilities,
are important for the successful explosion 
(\eg Blondin et al. 2003; Kotake et al. 2004; 
Buras et al. 2006; Burrows et al. 2006).

In these circumstances, observational constraints on the SN asymmetry 
are important.
Such observations include the direct imaging of 
Galactic young supernova remnants (\eg Hwang et al. 2004)
and SN 1987A (Wang et al. 2002),
although the number of such objects is limited.
The observations of extragalactic, point-source SNe also give clues
to the explosion geometry. 
For example, the shape of emission lines in optical spectra 
at $\gsim$ 1 year after the explosion can be used 
to study the multi-dimensional structure in
the innermost part of core-collapse SNe
(Mazzali et al. 2005; Maeda et al. 2008; Modjaz et al. 2008).

The most direct way to study the asymmetry of extragalactic, 
point-source SNe at early phases is polarimetry.
Since the polarized lights scattered by electrons in the ejecta are completely 
cancelled out in the spherically symmetric case,
the detection of polarization undoubtedly indicates the deviation 
from spherical symmetry (Shapiro \& Sutherland 1982; H\"oflich 1991).
With spectropolarimetry, the polarization across the P-Cygni profile
can give the information on the distribution of elements.
Spectropolarimetric studies have clarified 
the asymmetric nature of core-collapse SNe in detail 
(\eg Cropper et al. 1988; Trammell, Hines \& Wheeler 1993; Wang et al. 2001;
Leonard et al. 2001, 2006; Kawabata et al. 2002).

In this paper, we present the spectropolarimetric observation of SN 2007gr.
SN 2007gr was discovered in NGC 1058 (Li et al. 2007).
Thanks to the short distance to this host galaxy
\footnote{Crockett et al. (2008) adopted 10.6 Mpc for the distance to the SN
(Schmidt et al. 1992; Terry et al. 2002; Pilyugin et al. 2004)
while Valenti et al. (2008) adopted 9.3 Mpc (Silbermann et al. 1996).},
SN 2007gr reached 12.7 mag in the R band at maximum brightness 
(Valenti et al. 2008), 
making the spectropolarimetric observation possible.
Photometric and spectroscopic properties of SN 2007gr 
were shown by Valenti et al. (2008).
They classified this SN as Type Ic 
due to the absence of the H and He lines.
The progenitor of Type Ic SNe is thought to have lost 
its H and He layers before the explosion 
(\eg Wheeler et al. 1987; Nomoto et al. 1994).
Non-detection in the pre-explosion {\it Hubble Space Telescope} image 
at the SN position is consistent with a Wolf-Rayet star progenitor 
(Crockett et al. 2008).
The maximum brightness of SN 2007gr is similar 
to that of normal SNe, being powered by
the decay of $\sim$ 0.07$-$0.1 $\Msun$ \Nifs\ (Valenti et al. 2008).

We present in \S \ref{sec:obs}
the observation and data reduction of SN 2007gr.
Results of our spectropolarimetric observation are shown 
in \S \ref{sec:res}, where
the interstellar polarization (ISP) is also discussed.
We study the multi-dimensional explosion geometry 
of SN 2007gr in \S \ref{sec:str} 
and summarize conclusions in \S \ref{sec:con}.

\section{Observations and Data Reduction}
\label{sec:obs}

\begin{figure}
\begin{center}
\includegraphics[scale=0.48]{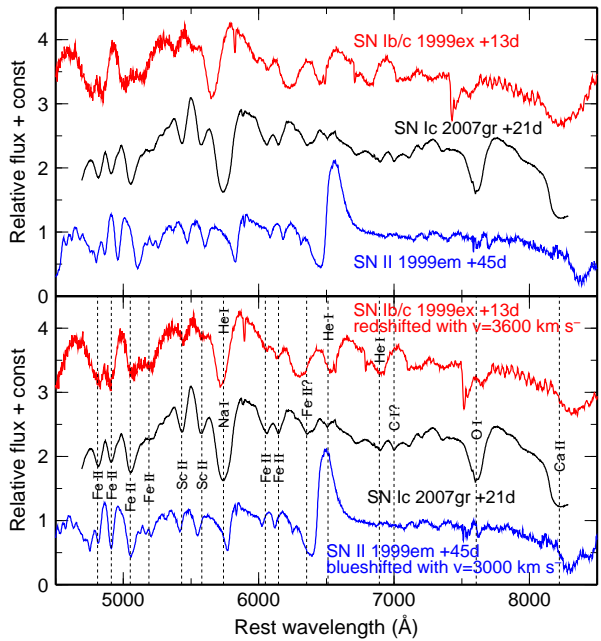}
\caption{
{\it Upper}: The spectrum of SN 2007gr (black) 
at 21 days after the maximum ($+21$ days), 
compared with the spectrum of SN Ib/c 1999ex 
at $+13$ days (red, Hamuy et al. 2002)
and of SN IIp 1999em at $+45$ days (blue, Leonard et al. 2001).
The redshift of the host galaxy is corrected.
{\it Lower}: Same as upper panel but the spectrum of SN 1999ex is 
redshifted by $v=3600$ \kms and the spectrum of SN 1999em is blueshifted
by $v=3000$ \kms.
\label{fig:spec}}
\end{center}
\end{figure}

The spectropolarimetric observation of SN 2007gr was performed 
with the 8.2 m Subaru telescope equipped with the 
Faint Object Camera and Spectrograph (FOCAS, Kashikawa et al. 2002)
on 2007 September 18 UT.
This epoch corresponds to 21 days after the maximum brightness,
which is $\sim 37$ days after the explosion 
assuming the rise time of 16 days (Valenti et al. 2008).

We used a $0\farcs 8$ width slit, a 300 lines mm$^{-1}$ grism 
and SY47 order-cut filter, giving a resolution of 
$\lambda/\Delta \lambda \sim 650$.
The linear polarimetric module of FOCAS consists of 
a rotating superachromatic half-wave plate 
and a crystal quartz Wollaston prism.
Both the ordinary and extraordinary rays are recorded 
on the CCD simultaneously.
With the four integrations
at $0^{\circ}, \ 45.^{\circ},\ 22.5^{\circ}$ and $67.5^{\circ}$ position 
of the half-wave plate, Stokes $Q$ and $U$ were derived as in
Tinbergen (1996).
For the degree of polarization $P$, the polarization bias correction 
was done using the results of Patat \& Romaniello (2006).
Total exposure time of four integrations was 1200 sec.

The flux was calibrated using the observation of the standard star 
G191B2B (Massey et al. 1988; Oke 1990).
Telluric absorptions in the SN spectrum were removed 
using those in the standard star spectrum.
Spectropolarimetry of this unpolarized star was also performed
with the same configuration as for the SN observation.
From this observation, it was confirmed
that the instrumental polarization is less than 0.1 \%.
The position angle was calibrated by the observation of dome flats
through the fully-polarizing filter with a known position angle.
The wavelength dependence of the optical axis of half-wave plate
was also corrected through this procedure.

To check the validity of the derived polarization spectrum,
we generated artificial shifts of $\pm$ 4 - 8 pix 
in the dispersion direction in one of four SN frames,
and did the same reduction.
This shift corresponds to the slit width (0$\farcs 8$ $\sim$ 8 pix) 
and this amount of shift could possibly happen within the slit width.
We confirmed that this shift does not alter the polarization features
discussed in this paper.

\begin{figure}
\begin{center}
\includegraphics[scale=0.45]{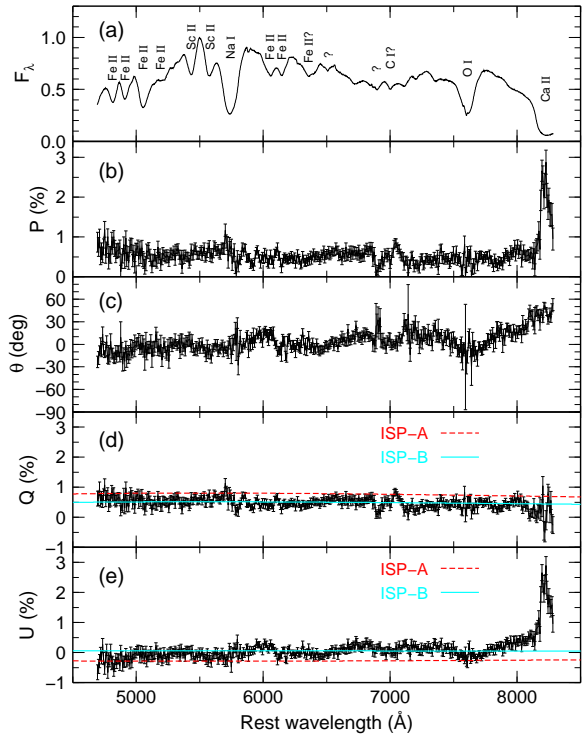}
\caption{
(a) The spectrum of SN 2007gr 
(in $10^{-14} {\rm erg\ s^{-1}\ cm^{-2}}$ \AA$^{-1}$) 
at 21 days after the maximum,
(b) the degree of polarization $P$, (c) polarization angle $\theta$, and  
(d, e) Stokes parameters $Q$ and $U$.
Interstellar polarization (ISP) 
is {\it not} corrected in $P$, $Q$, $U$ and $\theta$.
The polarization data are binned to 9 \AA, 
similar to the spectral resolution.
The red dashed line and cyan solid line show the 
two possible ISPs: 
($p_{\rm max, ISP\mathchar`-A}$, $\theta_{\rm ISP\mathchar`-A}$) 
= (0.80\%, 170$^{\circ}$) and 
($p_{\rm max, ISP\mathchar`-B}$, $\theta_{\rm ISP\mathchar`- B}$) 
= (0.51\%, 3$^{\circ}$), 
respectively.
\label{fig:pol}}
\end{center}
\end{figure}

\section{Results}
\label{sec:res}

\subsection{Spectroscopic Properties}

\begin{figure}
\begin{center}
\includegraphics[scale=0.45]{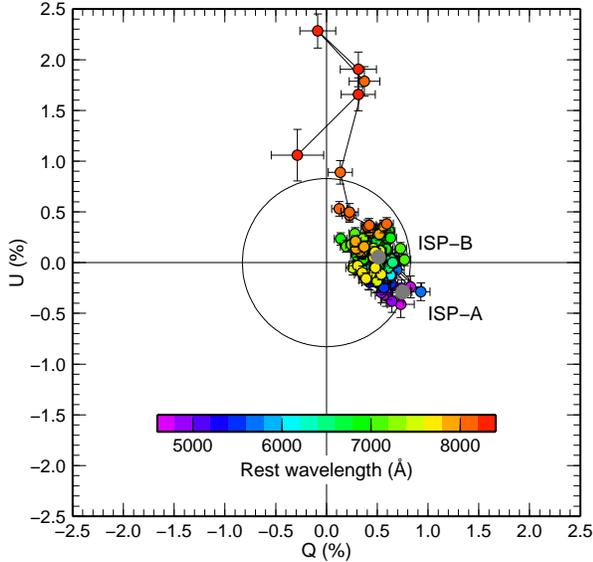}
\caption{
$Q$-$U$ diagram of the polarization data before 
the correction of interstellar polarization (ISP).
Different colors show the wavelength as shown in the color scale.
Possible two choices of ISP are marked with gray circles.
The polarization data are binned to 25 \AA.
The thin large circle shows the upper limit of ISP ($0.83 \%$).
\label{fig:qu}}
\end{center}
\end{figure}

The upper panel of Figure \ref{fig:spec} shows 
the spectrum of SN 2007gr (black) compared with
SN 1999ex at +13 days from B maximum (red, Hamuy et al. 2002) 
and SN 1999em at +45 days from B maximum (blue, Leonard et al. 2001).
These two SNe show most similar features to SN 2007gr at these epochs.
The later epoch of Type II SN 1999em reflects the slower evolution of 
Type II SNe than Type Ib/c SNe.
The redshift of the host galaxy is corrected.

If the spectrum of SN 1999ex is redshifted by $v=3600$ \kms\ (lower panel), 
the positions of several absorption minima in SN 2007gr 
are similar to those in the shifted spectrum of SN 1999ex, especially at 
the wavelength range redder than 6500 \AA.
Although there are weak troughs at the positions of the \ion{He}{i} 
$\lambda$6678 and 7065 in SN 2007gr,
these are unlikely to be the He lines because of
the absence of the \ion{He}{i} $\lambda$20581, which is usually stronger 
than the \ion{He}{i} lines in the optical range (Valenti et al. 2008).
Note that the absorption at 5750 \AA\ in SN 2007gr 
should be mainly by \ion{Na}{i} $\lambda$5889.

Valenti et al. (2008) noted that SN 2007gr shows 
the \ion{C}{i} and \ion{C}{ii}
lines in the optical and NIR wavelength range. 
In this respect, SN 2007gr is a special class of Type Ic SNe, 
or ``carbon-rich'' Type Ic SN (Valenti et al. 2008).
In our spectrum, 
one clear difference from the shifted spectrum of 
SN 1999ex is seen at 7000 \AA.
SN 2007gr has a weak absorption feature while SN 1999ex does not.
This could be the \ion{C}{i} $\lambda$7175 line.
The weak trough at 6500 \AA\ could be attributed to the \ion{C}{ii}
$\lambda$6578 line, but then the Doppler velocity of the line
($v \sim 3000$ \kms) is 
too small compared with other lines ($v \sim 7000$ \kms).

If the spectrum of SN 1999em is blueshifted by $v=3000$ \kms, 
the spectral features are in good agreement with those in SN 2007gr
especially at the wavelength range bluer than 6200 \AA.
These are mainly the absorptions of \ion{Na}{i}, 
\ion{Sc}{ii} and \ion{Fe}{ii} (Leonard et al. 2001).

\subsection{Spectropolarimetric Properties}

Figure \ref{fig:pol} shows the spectrum of SN 2007gr, 
the degree of polarization ($P \equiv \sqrt{Q^2 + U^2}$), 
polarization angle [$\theta \equiv 0.5 {\rm atan}(U/Q)$, 
measured from north to east], 
Stokes parameters $Q$ and $U$ (from upper to lower panels).
Note that the interstellar polarization (ISP) 
is {\it not} corrected in Figure \ref{fig:pol}.

The degree of polarization $P$ is $\sim 0.3 - 0.5$ \% 
at 4600 - 8000 \AA\ (Fig. \ref{fig:pol}).
Although the change in the degree of polarization 
at this wavelength range is not significant,
the polarization angle increases with the wavelength, caused
by the decreasing and increasing trend in Stokes $Q$ and $U$, respectively.
This does not directly imply that the SN has intrinsic
polarization with the increasing position angle with wavelength
since the data still suffer from ISP.

The properties of the spectropolarimetric data are more comprehensive
in the $Q$-$U$ diagram (Fig. \ref{fig:qu}).
In Figure \ref{fig:qu}, 
Stokes parameters at the different wavelength are shown 
in different colors, representing its wavelength as in the color scale below.
The majority of the data except for the reddest part
is aligned in straight line, although it is not very prominent.
The similar alignment in the $Q$-$U$ diagram
has been observed in many core-collapse 
SNe (\eg Wang et al. 2001; Leonard et al. 2002b).

The \ion{Ca}{ii} IR triplet has a large degree of polarization,
reaching $\sim 3$ \%.
The polarization is largest at the absorption minimum 
(Fig. \ref{fig:pol}).
The large polarization level of \ion{Ca}{ii} is independent 
on the choice of ISP (\S \ref{sec:ISP}).

No other line shows as strong polarization 
as in the \ion{Ca}{ii} line.
We see, however, a small peak of polarization at 5750 \AA\ 
(near the \ion{Na}{i} line)
and a fluctuation around 7000 \AA\ 
(near the possible \ion{C}{i} line).
Note that the intrinsic polarization of SN at these lines 
depends on the choice of ISP (\S \ref{sec:ISP}).

\begin{figure}[t]
\begin{center}
\includegraphics[scale=0.45]{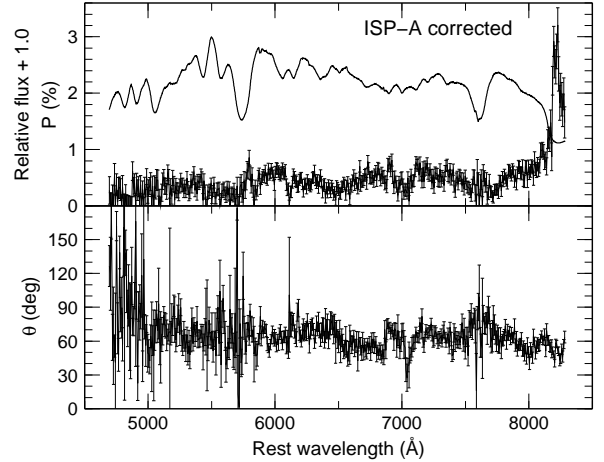}
\caption{
The polarization spectrum (upper) and position angle (lower) of SN 2007gr
corrected with ISP-A.
ISP-A is 
($p_{\rm max, ISP\mathchar`-A}$, $\theta_{\rm ISP\mathchar`-A}$) 
= (0.80\%, 170$^{\circ}$).
The polarization data are binned to 9 \AA.
\label{fig:ISPA}}
\end{center}
\end{figure}

\subsection{Interstellar and Intrinsic Polarization}
\label{sec:ISP}

In order to discuss the properties of SN,
the contribution of ISP in the line of sight to the SN position
should be removed from the observed data.
The amount of reddening gives the upper limit of ISP.
The total reddening is $E(B-V)_{\rm tot}=0.092$, 
which is the sum of the Galactic reddening $E(B-V)_{\rm Gal}=0.062$ and 
the host reddening $E(B-V)_{\rm host}=0.03$ (Valenti et al. 2008).
Then, the maximum degree of ISP is $p \sim 0.83$ \%, using 
the relation of $p /E(B-V) < 9 \%$ (Serkowski et al. 1975).
This maximum degree of ISP is shown in the circle in Figure \ref{fig:qu}.

To estimate the ISP, 
we should use SN data themselves unless detailed studies on ISP 
in this line of sight are available.
With SN data, ISP could be estimated by the following methods (1) -- (4);
(1) using polarimetry of SN at late epochs, when the SN ejecta 
become thin and electron scattering is not effective,
(2) following the time evolution of SN polarimetry
and assuming the time-independent component as ISP (\eg Leonard et al. 2002b),
(3) taking the polarization level at the strong P-Cygni emission peak,
where complete depolarization can be expected
(\eg Kawabata et al. 2002; Maund et al. 2007a), and 
(4) assuming SN photosphere has a single polarization angle, \ie 
an axisymmetric structure (\eg Wang et al. 2003b).

The methods (1) and (2) are rather robust, 
but unfortunately these cannot be used for our one-epoch data.
The method (3) can be taken even in one-epoch data.
However, since the emission peak of the most polarized \ion{Ca}{ii} IR triplet 
is not covered, this approach is also not effective for our data.
Although the method (4) relies on the SN theory,  
it can be applied to our SN 2007gr data because the polarization data
are aligned in the straight line in the $Q$-$U$ diagram (Fig. \ref{fig:qu}).
It should be emphasized, however, that this method is less reliable 
compared with the other three methods.

Adopting the method (4),
we could take both sides of the straight line in the $Q$-$U$ diagram 
as an estimated ISP.
Since it is expected that the bluer spectrum suffers
from more depolarization (\eg Howell et al. 2001; Wang et al. 2003b),
it is a sound assumption to take the bluest side as ISP.
The estimated ISP (ISP-A) is 
($p_{\rm max, ISP\mathchar`-A}$, $\theta_{\rm ISP\mathchar`-A}$) 
= (0.80\%, 170$^{\circ}$)
as shown in the gray circle in Figure \ref{fig:qu}.
Here, $p_{\rm max}$ is the degree of ISP at $\lambda_{\rm max}$.
We assume $\lambda_{\rm max}= 5500$\AA, 
which is the wavelength of the maximum ISP (Serkowski et al. 1975).
The wavelength dependence of ISP is taken from
Serkowski et al. (1975);
\begin{equation}
p(\lambda)\ 
=\ p_{\rm max}\exp \left[ 
-K\ {\rm ln}^2 \left( \frac{\lambda_{\rm max}}{\lambda} \right) \right],
\end{equation}
which is commonly accepted for ISP in our galaxy.
Here $K$ is given as 
$K = 0.01 + 1.66\ \lambda_{\rm max}\ ({\rm \mu m}) =0.932$
(Whittet et al. 1992).
The $Q$ and $U$ components of ISP-A are shown in the red dashed line
in Figure \ref{fig:pol}
\footnote{
It should be noted that the wavelength dependence of ISP in the host 
galaxy could be different from Serkowski's law.
If the same wavelength dependence can be assumed,
it is not a bad approximation to use the wavelength dependence of 
Serkowski's law for the sum of ISPs in our Galaxy and the host galaxy
because ISP-A is close to the maximum degree of ISP.
}.

\begin{figure}[t]
\begin{center}
\includegraphics[scale=0.45]{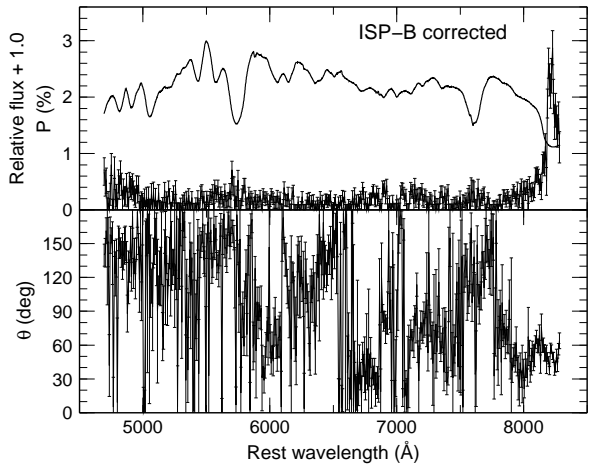}
\caption{
Same as Fig. \ref{fig:ISPA}, but polarization data
are corrected assuming ISP-B
($p_{\rm max, ISP\mathchar`-B}$, $\theta_{\rm ISP\mathchar`-B}$) 
= (0.51\%, 3$^{\circ}$).
The spectra are binned to 9\AA.
\label{fig:ISPB}}
\end{center}
\end{figure}

The polarization data after the subtraction of ISP-A are
shown in Figure \ref{fig:ISPA}.
The polarization angle ($\theta$) is $\sim 40 - 80^{\circ}$ for
a wide range of the spectrum,
although there is a large scatter in the blue wavelength.
The polarization angle in the line-free region at 6600 - 6700 \AA\
is $60 \pm 5^{\circ}$.
The degree of polarization is $0.3 - 0.7 \%$ except for the reddest part,
and it is $\sim 0.5 \pm 0.2 \%$ at 6600 - 6700 \AA.

It should be noted that 
there are cautions when we take ISP-A.
Since the reddening neither in our Galaxy nor the host galaxy is 
negligible (Valenti et al. 2008), 
it is expected that the light from SN suffers from ISP 
both in our Galaxy ($\lsim$ 0.56\%) and 
the host galaxy ($\lsim$ 0.27 \%).
Thus, the choice of ISP-A ($p_{\rm max, ISP\mathchar`-A} = 0.8$ \%), being
close to the upper limit of ISP ($p_{\rm max}=0.83$ \%), 
requires a fine alignment of the position angles of ISP,
which could be unlikely.
In addition, although 
the position angle of ISP in the host galaxy is expected to 
be aligned with the spiral arm ($\theta \sim 30^{\circ}$ in this case,
see \eg Scarrott, Ward-Thompson, \& Warren-Smith 1987; Scarrott et al. 1993), 
the position angle of ISP-A is displaced by $\sim 40^{\circ}$ 
from that of the spiral arm.

As an alternative choice of ISP, we take ISP-B at the 
center of polarization data in the $Q$-$U$ diagram,
($p_{\rm max, ISP\mathchar`-B}$, $\theta_{\rm ISP\mathchar`-B}$) 
= (0.51\%, 3$^{\circ}$),
which is shown in the gray circle in Figure \ref{fig:qu}.
The $Q$ and $U$ components of this ISP are shown in Figure \ref{fig:pol}
in the cyan solid line 
\footnote{We simply assume Serkowski's law also in ISP-B, 
meaning the alignment of position angle of ISP in our Galaxy and host galaxy,
which is not necessarily required.
But the wavelength dependence of ISP is very weak and this treatment 
is essentially similar to the subtraction of a wavelength-independent ISP.}.
Figure \ref{fig:ISPB} shows the polarization after the correction of ISP-B.
Since the straight line in the $Q$-$U$ diagram is not very prominent,
this choice results in the small degree of intrinsic polarization 
at a wide wavelength range except for the \ion{Ca}{ii} IR triplet 
(upper panel).
Although the polarization angle is very scattered 
due to the small $Q$ and $U$, 
a large portion of the data has
$\theta \sim 150^{\circ}$ and $\sim 60^{\circ}$ 
at mainly blue and red wavelength range, respectively. 

Both ISP-A and ISP-B are not inconsistent with the 
polarization degrees and angles of field stars
in the stellar polarization catalog of Heiles (2000) 
close to the SN position.
Thus, with the available data only, 
we cannot distinguish the two possibilities of ISP.
We thus discuss possible interpretations of the polarization data
with these two ISPs.

\begin{figure}
\begin{center}
\includegraphics[scale=0.62]{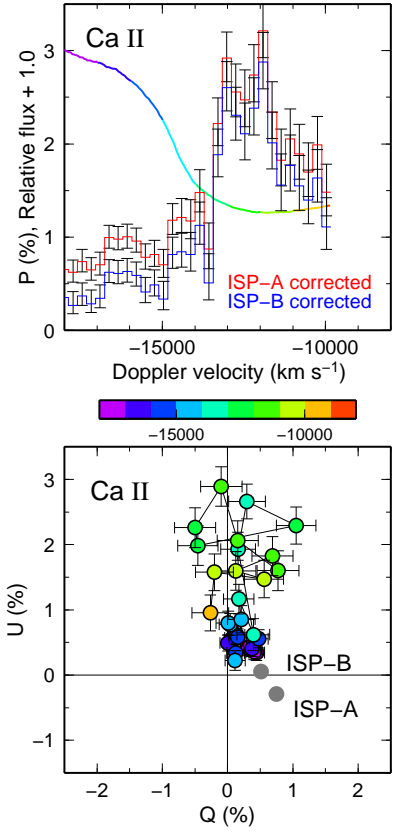}
\caption{
Polarization properties across 
the \ion{Ca}{ii} IR triplet (mean $\lambda$8567) line.
{\it Upper panels}: The spectrum (color line) and 
the degree of the polarization.
The red and blue lines show the polarization 
corrected with ISP-A and ISP-B, respectively.
They are shown against the Doppler velocity
measured from the rest wavelength.
{\it Lower panels}: {\it ISP-uncorrected} $Q$-$U$ diagrams 
of the \ion{Ca}{ii} line.
The ISP-A and ISP-B are shown in by the gray points.
The color of each point shows the Doppler velocity.
The spectra are binned to 9\AA.
\label{fig:lines1}}
\end{center}
\end{figure}

\section{Geometry of SN 2007\lowercase{gr}}
\label{sec:str}

In the following sections, we discuss the multi-dimensional 
ejecta structure of SN 2007gr.
With the continuum polarization, we can infer the asymmetry of 
electron scattering-dominated photosphere 
(\eg Shapiro \& Sutherland 1982; H\"oflich 1991).
In addition, with the polarization properties across the P-Cygni profiles, 
we can discuss the distribution of elements.
In strong P-Cygni profiles, the polarization often has maximum (minimum) at 
the minimum (maximum) of the flux,
which is so called ``inverted P-Cygni'' profile (Jeffery 1989, 1991).
The polarization level becomes maximum at the absorption minimum
because the lights that are scattered through smallest angles by electrons
(and thus have less polarization) are effectively scattered out 
by the line at the absorption minimum.
The polarization minimum at the emission peak is caused by the dilution
due to the directly-coming, unpolarized emission by the line.

In a simple ellipsoidal case, 
where only the unpolarized or polarization-cancelled lights 
are scattered by the line,
the polarization at the absorption trough ($p_{\rm trough}$) is limited as
\begin{equation}
p_{\rm trough} \lsim p_{\rm cont} \frac{I_{\rm cont}}{I_{\rm trough}},
\label{eq:pol}
\end{equation}
where $p_{\rm cont}$, $I_{\rm cont}$, and $I_{\rm trough}$ are 
the continuum polarization, the continuum flux, and 
the flux at the trough, respectively (Leonard et al. 2001).
If the line polarization larger than the limit of Eq. (\ref{eq:pol}) 
is observed at the trough, 
it means that the simple ellipsoidal model does not work
and that the clumpy or patchy line forming region obscures 
a certain part of the photosphere, enhancing the polarization level.

\begin{figure*}
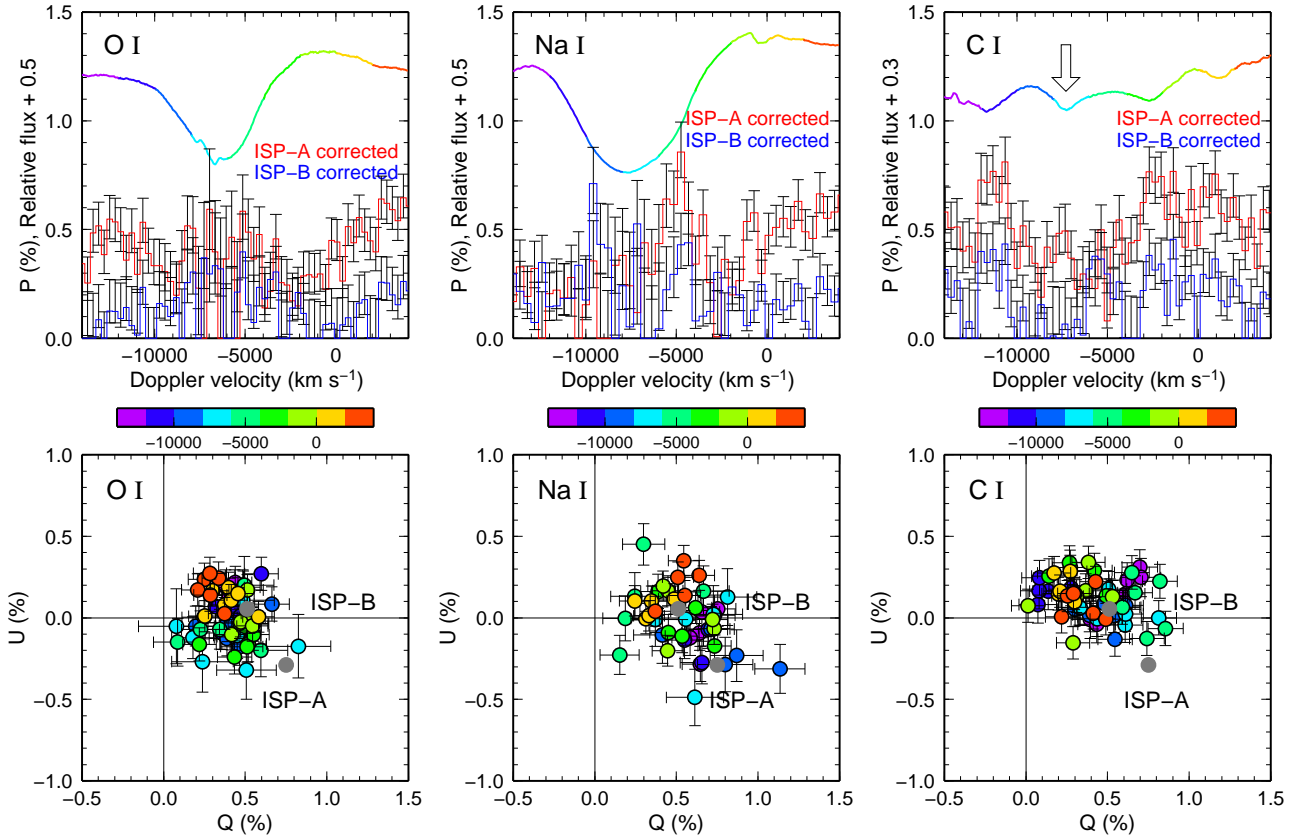

\begin{center}
\begin{tabular}{ccc}
\includegraphics[scale=0.62]{f7a.eps}&
\includegraphics[scale=0.62]{f7b.eps}&
\includegraphics[scale=0.62]{f7c.eps}
\end{tabular}
\caption{
Polarization properties across 
the \ion{O}{i} $\lambda$7774, \ion{Na}{i} $\lambda$5889,
and possible \ion{C}{i} $\lambda$7175 lines (from left to right).
{\it Upper panels}: The spectrum (color line) and 
the degree of the polarization.
The red and blue lines show the polarization 
corrected with ISP-A and ISP-B, respectively.
They are shown against the Doppler velocity
measured from the rest wavelength.
{\it Lower panels}: {\it ISP-uncorrected} $Q$-$U$ diagrams 
of the \ion{Ca}{ii} line.
The ISP-A and ISP-B are shown in by the gray points.
The color of each point shows the Doppler velocity.
The spectra are binned to 9\AA.
\label{fig:lines2}}
\end{center}
\end{figure*}

In the upper panel of Figure \ref{fig:lines1},
the flux spectrum and the ISP-corrected polarization spectrum 
versus Doppler velocity are shown near the 
\ion{Ca}{ii} IR triplet (mean $\lambda$8567).
Figure \ref{fig:lines2} is
the same as Figure \ref{fig:lines1} but for 
the \ion{O}{i} $\lambda$7774, \ion{Na}{i} $\lambda$5889,
possible \ion{C}{i} $\lambda$7175 lines.
The red and blue lines show the ISP-A and ISP-B corrected 
polarization, respectively.
In the lower panels of Figures \ref{fig:lines1} and \ref{fig:lines2},
the $Q$-$U$ diagrams at these lines are shown.
Since the change of ISP across the line is negligible,
ISP is {\it not} corrected for in the $Q$-$U$ diagram, where
the two choices of ISP are shown by the gray circles.
The color in these plots shows the Doppler velocity.
The spectra in Figures \ref{fig:lines1} and \ref{fig:lines2} 
are binned to 9\AA.

\begin{figure}
\begin{center}
\includegraphics[scale=0.4]{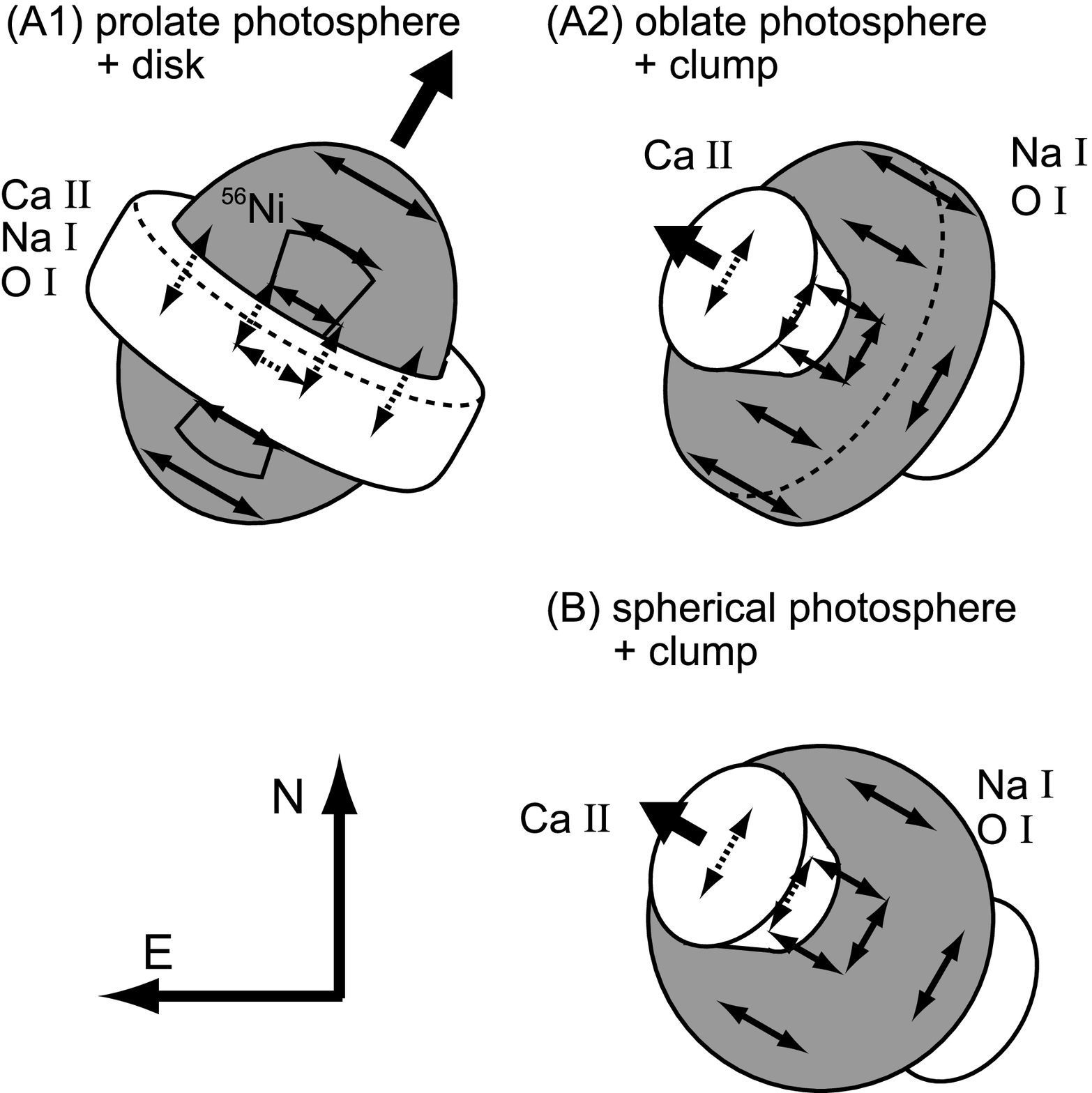}
\caption{
Schematic illustrations of possible bipolar explosion models.
Shaded area shows the photosphere.
The thin arrows represent the polarization vector. 
Photons scattered-out by the line forming region 
(unshaded area) are shown in dashed arrows.
The thick arrows show the polar direction.
A1 and A2 are the models with ISP-A while B is with ISP-B.
The degree of asphericity is exaggerated.
Both A1 and A2 would show continuum polarization with 
the position angle of $\sim 60^{\circ}$. 
The strong line polarization at the \ion{Ca}{ii} 
line with similar position angle can also be explained.
However, case A1 predicts similarly strong line polarization 
at the \ion{O}{i} and \ion{Na}{i} lines.
For case A2, \ion{O}{i} and \ion{Na}{i} may follow the photosphere,
which is consistent with the observed polarization.
Case B does not have continuum polarization but has 
line polarization at the \ion{Ca}{ii} line with the position angle of 
$\sim 60^{\circ}$.
\label{fig:model}}
\end{center}
\end{figure}

\subsection{Interstellar Polarization-A: Axisymmetric Photosphere}
\label{sec:ISPA}

Adopting ISP-A, the intrinsic SN polarization has
an almost constant position angle (Fig. \ref{fig:ISPA}).
If we take the line free region (6600-6700 \AA) as the continuum region,
the continuum has the polarization of $\sim 0.5 \%$, with 
the position angle $\theta \sim 60^{\circ}$.
This suggests that the shape of the SN photosphere projected to the sky 
is deformed at a level of $\sim 10$ \% (\eg H\"oflich 1991).
The major axis of the ellipsoid is located in the direction of 
$\sim 150^{\circ}$ measured from north to east 
(\ie $\sim 30^{\circ}$ from north to west)
because the polarization is orthogonal to the radial direction 
of the projected photosphere (see A1 and A2 Fig. \ref{fig:model}).

\subsubsection{Calcium}

The polarization across the \ion{Ca}{ii} line is prominent,
reaching $\sim 3$\% at the absorption minimum
as seen from the red line in the upper left panel of Figure \ref{fig:lines1}.
The polarization angle of the \ion{Ca}{ii} line is $\sim 50^{\circ}$, 
which is close to that of the continuum ($\sim 60^{\circ}$).

At the \ion{Ca}{ii} line in SN 2007gr, 
$I_{\rm cont}/I_{\rm trough} \sim 5-6.5$
given the uncertainty of the continuum level.
Taking the continuum polarization is $p_{\rm cont} \sim 0.5 \%$,
Eq. (\ref{eq:pol}) gives the limit of $p \lsim 2.5-3.25 \%$.
Thus, although the simple ellipsoidal model could be accepted,
a selective obscuration enhancing the uncancelled polarization component
is preferred to robustly explain the observed high polarization 
level at the line ($\sim 3\%$),
considering the idealized situation assumed in Eq. (\ref{eq:pol}).
Since the position angle of the \ion{Ca}{ii} line is similar to that of the 
continuum, the \ion{Ca}{ii} line forming region should selectively obscure 
the light originated around the minor axis of the SN photosphere
(see A1 and A2 of Fig. \ref{fig:model}).

\subsubsection{Oxygen, Sodium and Carbon}

Although the absorption line of \ion{O}{i} is strong, 
no significant change of polarization is 
observed across the line (the left panel of Fig. \ref{fig:lines2}).
At this line, $I_{\rm cont}/I_{\rm trough}\sim 2$, 
and thus, polarization of $\sim 1\%$ 
at the absorption minimum can be expected from Eq. (\ref{eq:pol}).
This means that the \ion{O}{i} line obscures not only 
unpolarized or polarization-cancelled light but also 
some part of the projected photosphere near its major axis.
Thus, the distribution of \ion{O}{i} follows the photosphere, 
being clearly different from that of \ion{Ca}{ii}.

The polarization at the \ion{Na}{i} line is marginally enhanced
(the center panel in Fig \ref{fig:lines1}),
although the maximum of polarization is not located exactly 
at the absorption minimum of the flux.
Note that this behavior depends on the choice of ISP.
Overall, the line does not have significant polarization as expected from 
Eq. (\ref{eq:pol}), suggesting a similar distribution of \ion{Na}{i} 
to \ion{O}{i}.

There is no large change of the polarization 
at the possible \ion{C}{i} line (the right panel of Fig. \ref{fig:lines2}).
However, this is just a natural consequence of the weakness of the line
(\ie $I_{\rm cont}/I_{\rm trough} \sim 1$).

\subsubsection{Geometry of the Ejecta}

We summarize the polarization properties under the assumption of ISP-A.
The continuum polarization suggests that
there is a mild asphericity in the photosphere.
The distribution of \ion{O}{i} and \ion{Na}{i}  
follows the photosphere.
On the other hand, the distribution of \ion{Ca}{ii} is different, 
which is likely to obscure the photosphere near its minor axis.
This effectively enhances the polarization level, 
unchanging the polarization angle.
There is not much information on the distribution of \ion{C}{i}.

If the photosphere has a prolate shape, 
the observed properties might be consistent with
the bipolar explosion seen from near the equatorial plane
(A1 of Fig. \ref{fig:model}).
The pole points to $\sim 30^{\circ}$ from north to west in the sky.
In the bipolar explosion (\eg Khokhlov et al. 1999; Maeda et al. 2002), 
the heavy element including \Nifs\ is synthesized in the polar region.
In such a model, the polar region is more ionized and 
a prolate photosphere can be formed
as suggested by H\"oflich (2001) for Type II SNe.
Further, a disk-like structure remains in the equatorial plane
and \ion{Ca}{ii} can also be enhanced in the disk 
due to the low temperature (Tanaka et al. 2007).
However, in this scenario, the distribution of \ion{Na}{i} and \ion{O}{i} 
may follow \ion{Ca}{ii}, which is not consistent with the observation.

Alternatively, in the bipolar explosion, the photosphere could 
have an oblate shape, tracing the dense disk-like region.
This may be true at $\gsim 30$ days after the explosion (Maeda et al. 2006).
Then the observed continuum polarization requires that 
the explosion is seen from the slightly off-axis direction 
(A2 of Fig. \ref{fig:model}).
In this case, explosively synthesized Ca
near the polar region can make the \ion{Ca}{ii} absorption.
In addition, the distribution of \ion{O}{i} and \ion{Na}{i} may follow
the photosphere (Tanaka et al. 2007).
This ejecta structure is qualitatively consistent with the observed properties.

\subsection{Interstellar Polarization-B: Almost Spherical Photosphere}
\label{sec:ISPB}

\begin{deluxetable*}{lccccccc} 
\tablewidth{0pt}
\tablecaption{Comparison of Spectropolarimetric Features}
\tablehead{
SN name $^{\rm Ref.}$ &
Type &
Epoch \tablenotemark{a} &
$p_{\rm cont}$ \tablenotemark{b} & 
$\Delta p_{\rm Ca}$ \tablenotemark{c} &
$\Delta p_{\rm O}$ \tablenotemark{d} &
$\Delta p_{\rm Na}$ \tablenotemark{e} &
$\Delta p_{\rm Fe}$ \tablenotemark{f} 
}
\startdata
SN 2007gr (ISP-A)  & Ic       & 37    & $\sim$ 0.5   & 
2.5   &  0 &  0  &  0  \\
SN 2007gr (ISP-B)  &          &       & $\lsim$ 0.2  &
2.5   &  0 &  0  &   0 \\
\hline
SN 2002ap $^{1}$   & broad Ic & 11-16 &  $\sim $ 0.5  &
$\lsim$ 0.2  & 0.4  & - & -  \\ 
                         &          & 37-41  &  $\sim$ 0.5-1   &
 1.5  &  0.2  & - & - \\ 
SN 2005bf $^{2}$               & pec. Ib & 34  & $\lsim$ 0.45 &
3.5  & -  & -  & 1  \\ 
SN 2003gf $^{3}$             & Ic       & 30  & $\sim$ 0.5  & 
1.5   &  1.3  & 0.4  & 0.5 (?) \\ 
SN 2004dj $^{4}$               &  IIp    & 39   &  0   &
0        &   0      &    0       &     0       \\ 
                                &        & 91   &  0.6 &
 0       &  -         &    0.7   &   0   \\
SN 1987A  $^{5}$               &  pec. II  &   100    &  $\sim$ 0.5  &
3   &      -       &    $-$0.2   &    -       \\
\enddata
\tablenotetext{a}{Estimated days after the explosion} 
\tablenotetext{b}{Continuum polarization (\%)} 
\tablenotetext{c}{Polarization change at \ion{Ca}{ii} line trough (\%)} 
\tablenotetext{d}{Polarization change at \ion{O}{i} line trough (\%)}
\tablenotetext{e}{Polarization change at \ion{Na}{i} line trough (\%)}
\tablenotetext{f}{Polarization change at \ion{Fe}{ii} line trough (\%)}
\tablerefs{(1) Kawabata et al. (2002); Leonard et al. (2002b);
Wang et al. (2003b), (2) Maund et al. (2007), (3)
Leonard \& Filippenko (2005), (4) Leonard et al. (2006), 
(5) Cropper et al. (1988); Jeffery (1991)}
\label{tab:comp}
\end{deluxetable*}

If we take ISP-B, the polarization is 
very small at a wide wavelength range (Fig. \ref{fig:ISPB}).
The photosphere has a possible two-axis structure 
with the position angles of $\sim 150^{\circ}$ and $60^{\circ}$ 
at the blue and red spectra, respectively, although it is quite marginal
given the small polarization level (lower panel of Fig. \ref{fig:ISPB}).

Small continuum polarization suggests an almost spherical photosphere.
The small continuum polarization could be realized more easily 
when the radial density profile is steeper because of the smaller
contribution of the orthogonally-scattered, and thus
more-polarized lights.
This is consistent with the suggestion by Valenti et al. (2008):
the steep density profile is required to explain 
the slow recession of the photosphere.

If there is little continuum polarization, from Eq. (\ref{eq:pol}), 
any large line polarization undoubtedly 
suggests an aspherical distribution of the ion,
which can cause the selective obscuration of the photosphere.
This is similar to what is usually found in Type Ia SNe 
(\eg Wang et al. 2003a; Leonard et al. 2005b).

In fact, with the assumption of ISP-B, 
a strong polarization is seen in the \ion{Ca}{ii} line.
This requires an asymmetric distribution of \ion{Ca}{ii},
and the clump(s) or disk-like distribution may be possible.

There is no conspicuous polarization change across 
the \ion{O}{i}, \ion{Na}{i} and \ion{C}{i} line.
This suggests that these elements are distributed almost 
spherically, similar to the photosphere.

In summary, the SN polarization with ISP-B can be explained 
by the almost spherical photosphere.
Only the distribution of \ion{Ca}{ii} is largely 
deviated from spherical symmetry, and 
other ions such as \ion{O}{i}, \ion{Na}{i}, and \ion{C}{i}
are distributed almost spherically.

Constraints on the explosion models are not strong.
In the bipolar explosion models, 
Ca in the pre-explosion star could be enhanced 
in disk-like region (see A1 of Fig. \ref{fig:model}) 
due to the high density and the low ionization there.
However, as discussed in \S \ref{sec:ISPA},
this may not be consistent with the observed properties 
because O and Na are also abundant before the explosion, \ie
\ion{O}{i} and \ion{Na}{i} may also have the disk like distribution.

Alternatively, the explosively synthesized \ion{Ca}{ii} 
distributed near the axis (B in Fig. \ref{fig:model})
can explain the observation.
This may be realized by less-aspherical explosion than 
case A2 in Figure \ref{fig:model}.

\subsection{Comparison with Other Core-Collapse SNe}

There are only a small number of 
spectropolarimetric observations of Type Ib/c SNe to date
(SN Ic 1997X; Wang et al. 2004, SN Ic 1997ef; Wang et al. 2004,
SN Ic 1998bw; Patat et al. 2001, 
SN Ic 2002ap; Kawabata et al. 2002; Leonard et al. 2002b; Wang et al. 2003b, 
SN Ic 2003dh; Kawabata et al. 2003, 
SN Ic 2003gf; Leonard \& Filippenko 2005a,
SN Ib 2005bf; Maund et al. 2007a; 
SN Ic 2006aj; Maund et al. 2007b).
Note that SNe 1997ef, 1998bw, 2002ap, 2003dh, 2006aj are all broad-line 
Type Ic SNe, among which SNe 1998bw and 2003dh are estimated to have 
an explosion energy exceeding $10^{52}$ erg, thus being called 
hypernovae (\eg Nomoto et al. 2007).
Thus, SN 2007gr is one of the rare spectropolarimetric sample of 
normal Type Ib/c SNe.

We first compare the results with SN 2002ap, 
best observed Type Ic SN in spectropolarimetry
(Kawabata et al. 2002; Leonard et al. 2002b; Wang et al. 2003b).
There are also fair similarities between ISP-A subtracted 
SN 2007gr at 21 days after the maximum
($\sim 37$ days after the explosion)
and SN 2002ap at $30$ days after the maximum
($\sim 41$ days after the explosion, assuming the rise time of 11 days,
Mazzali et al. 2002):
(1) a moderate continuum polarization level ($\sim 0.5 \%$),
and (2) a strong polarization in the \ion{Ca}{ii} IR triplet
with the similar position angle to that of the continuum 
(see March 2002 data in Kawabata et al. 2002; Leonard et al. 2002b).
Note that SN 2002ap shows the two distinct polarization angle in
continuum and the Ca-O absorption at maximum epoch.

The light curve of SN 2007gr is similar to 
SN 2002ap but the spectra of SN 2007gr show narrower lines than in SN 2002ap
(Valenti et al. 2008).
Valenti et al. (2008) speculated that SN 2007gr
and SN 2002ap are the bipolar explosions observed from the minor and 
major axis, respectively.
Our data do not support this hypothesis
since there are similarities in polarization data of two objects and 
the polarization of SN 2007gr is likely to be interpreted by the 
viewing angle near the major axis (\S \ref{sec:ISPA} and \S \ref{sec:ISPB}).

Another comparison can be made with the peculiar SN Ib 2005bf
(Anupama et al. 2005; Tominaga et al. 2005; Folatelli et al. 2006; 
Maeda et al. 2007).
Maund et al. (2007) showed spectropolarimetric data of SN 2005bf at 
6 days before the 2nd maximum, which is $\sim 34$ days after the explosion
(assuming the rise time of $\sim$ 40 days, Tominaga et al. 2005).
SN 2005bf also showed a continuum polarization ($\sim 0.45\%$) 
and a large polarization level across the \ion{Ca}{ii} 
triplet with the polarization angle almost aligned to 
that of the photosphere.
In SN 2005bf, however, there are distinct polarization features 
at the \ion{Fe}{ii} lines, which are not seen in SN 2007gr.
This may be due to the difference in the degree of asphericity, 
the line of sight, or the epoch of the observation, 
\ie difference in the observed layer.

The comparison with Type II SNe after the plateau phase is also useful.
In Type II SNe, the increase of polarization degree 
after the plateau phase have been observed (\eg Leonard et al. 2001, 2006).
At this epoch, the He-core is exposed, 
which would be a similar to an exploding He/C+O star
(\ie Type Ib/c SNe).

In SN 2004dj (Leonard et al. 2006), the continuum polarization is 
$\sim 0.6 \%$ after the plateau phase, which is similar to that of
SN 2007gr.
Large line polarization changes are observed at the \ion{Na}{i}, H$\alpha$
and \ion{Ca}{ii} lines.
At the latter two lines, the P-Cygni emission component is much stronger
than the absorption and only the depolarization is observed.
At the \ion{Na}{i} line, polarization change at a level of $\sim 0.7 \%$
is observed, which is not seen in SN 2007gr.
In SN 1987A, on the other hand, 
a small depolarization at the \ion{Na}{i} line and 
a large polarization change $\sim 3 \%$ at the \ion{Ca}{ii} IR triplet
are observed (Cropper et al. 1988; Jeffery 1991).

The polarization properties discussed above are summarized 
in Table \ref{tab:comp}

\section{Conclusions}
\label{sec:con}

We have presented optical spectropolarimetric observations 
of Type Ic SN 2007gr at 21 days after maximum brightness
($\sim 37$ days after the explosion).
A strong polarization feature is observed across the \ion{Ca}{ii} 
IR triplet line, which is as high as $\sim 3$ \% independently of 
the choice of ISP.

If we assume that the SN photosphere has a single polarization angle,
the intrinsic polarization of the SN continuum is $\sim 0.5$\%.
This suggests that the axis ratio of the photosphere projected on the sky 
is different from unity by $\sim 10$\%.
The polarization angle of the continuum is $\sim 60^{\circ}$,
which is similar to that of the strong polarization in the 
\ion{Ca}{ii} line ($\sim 50^{\circ}$).
The polarization of the continuum and the \ion{Ca}{ii} line 
in SN 2007gr are similar to 
those found in broad-line SN Ic 2002ap after maximum
but not similar to SN 2002ap around maximum.

The spectropolarimetric features of SN 2007gr can be explained 
by a bipolar explosion viewed from the slightly off-axis direction
(A2 in Fig. \ref{fig:model}).
The photosphere has an oblate shape in the equatorial plane.
The explosively synthesized \ion{Ca}{ii} is distributed near the 
polar region, and it partially obscure the photosphere.
The distributions of \ion{O}{i} and \ion{Na}{i} 
follow the photosphere, being different from \ion{Ca}{ii}.
Given the uncertainty of ISP, however, the polarization data 
could also be interpreted by almost spherically symmetric photosphere 
and the aspherical \ion{Ca}{ii} distribution (B in Fig. \ref{fig:model}).

\acknowledgments

We would like to thank the anonymous referee for valuable comments.
M. T. would like to thank G. C. Anupama, D. K. Sahu, Daisuke Kato,
and Masafumi Shimada for helpful suggestions and discussions.
M. T. is supported by the JSPS (Japan Society for the Promotion of Science) 
Research Fellowship for Young Scientists.
This research has been supported in part by World Premier
International Research Center Initiative (WPI Initiative), MEXT,
Japan, and by the Grant-in-Aid for Scientific Research of the JSPS
(18104003, 18540231, 20540226) and MEXT (19047004, 20040004).

\end{document}